\documentclass[onecolumn,superscriptaddress,showpacs,pra]{revtex4}

\usepackage[dvips]{graphicx,color}
\usepackage{delarray}
\usepackage{amsmath}
\usepackage{bm}

\newcommand{\ket}[1]{| { #1} \rangle}
\newcommand{\bra}[1]{ \langle {#1}  |}

\begin{document}
\title{Fundamental rate-loss tradeoff for the quantum internet}

\author{Koji Azuma}
\email{azuma.koji@lab.ntt.co.jp}
\affiliation{NTT Basic Research Laboratories, NTT Corporation, 3-1 Morinosato Wakamiya, Atsugi, Kanagawa 243-0198, Japan}

\author{Akihiro Mizutani}
\affiliation{Department of Materials Engineering
Science,
Graduate School of Engineering Science, 
Osaka University, 1-3 Machikaneyama, Toyonaka, Osaka 560-8531, Japan}

\author{Hoi-Kwong Lo}
\affiliation{Department of Physics and Department of Electrical \& Computer Engineering,
University of Toronto, Toronto, Ontario, Canada}

\
\date{\today}

\begin{abstract}
The quantum internet holds promise for performing quantum communication---such as quantum teleportation and quantum key distribution (QKD)---freely between  any parties all over the globe. Such a future quantum network, depending on the communication distance of the requesting parties, necessitates to invoke several classes of optical quantum communication such as point-to-point communication protocols, intercity QKD protocols and quantum repeater protocols.
Recently, Takeoka, Guha and Wilde (TGW) have presented a fundamental rate-loss tradeoff on quantum communication capacity and secret key agreement capacity of any lossy channel assisted by unlimited
forward and backward classical communication [Nat.~Commun.~{\bf 5}, 5235 (2014)]. However, this bound is applicable only to the simplest class of quantum communication, i.e., the point-to-point communication protocols, and it has thus remained open to grasp the potential of a `worldwide' quantum network.
Here we generalize the TGW bound to be applicable to any type of two-party quantum communication over the quantum internet, including other indispensable but much more intricate classes of quantum communication---intercity QKD protocols and quantum repeater protocols.
We also show that there is essentially no scaling gap between our bound and the quantum communication efficiencies of known protocols. Therefore, our result---corresponding to a fundamental and practical limitation for the quantum internet---will contribute to design an efficient quantum internet in the future.
\end{abstract}
\maketitle

\begin{figure}
\includegraphics[keepaspectratio=true,height=75mm]{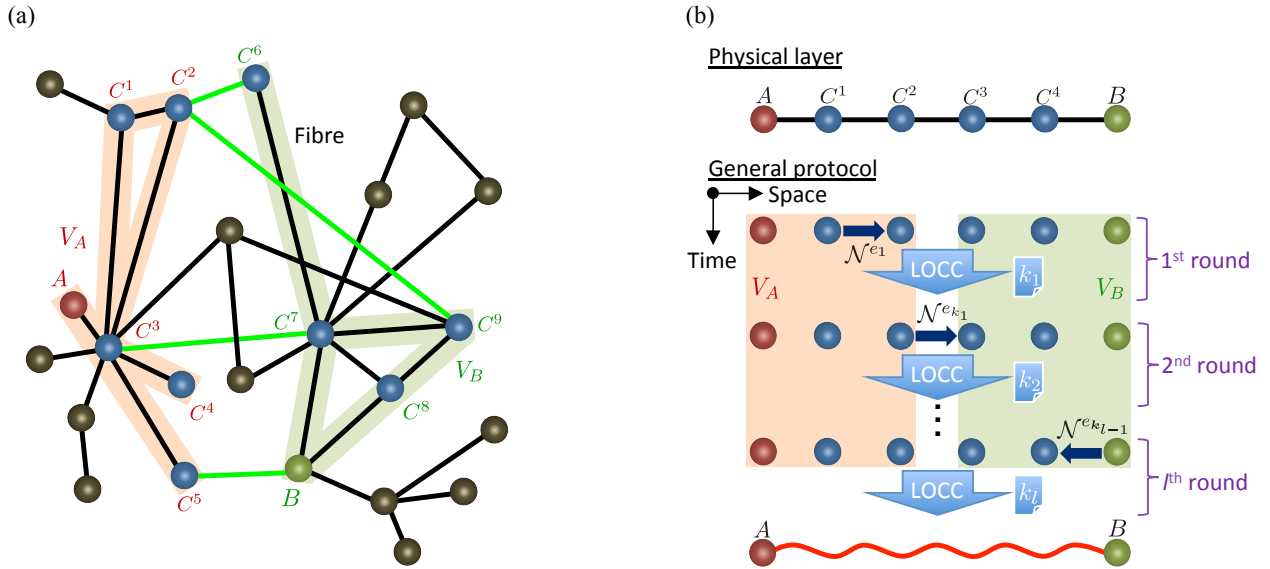}
  \caption{Quantum internet and the most general protocol. 
Panel (a) depicts a general quantum internet where Alice and Bob request its internet protocol to supply them with the resources for quantum communication, such as unconditionally secure key and quantum entanglement. Accordingly, the quantum internet protocol considers any quantum network $G$ (which might be a quantum subnetwork of a quantum internet) associated with a directed graph $G=(V,E)$.
The set $V$ of vertices is composed of the nodes as $V=\{A,B,C^1,C^2,\ldots,C^{n}\}$ ($n=9$ in this panel) and the set $E$ of edges specifies quantum channels $\{{\cal N}^{e} \}_{e \in E}$ in such a way that a quantum channel to send a quantum system from node $v_1 \in V$ to node $v_2\in V$ is represented by ${\cal N}^{v_1 \to v_2} $. The quantum internet protocol can combine given quantum channels $\{{\cal N}^{e} \}_{e \in E}$ with LOCC arbitrarily, to provide the required resources for Alice and Bob.  However, our bound suggests that
the obtainable secret bits or ebits are upper bounded by a bound for the point-to-point communication between a single parity having nodes $V_A \subset V$ with $A$ and another party having $V_B(=V\setminus V_A)$ with $B$. In panel (b), we describe the paradigm of the most general two-party communication protocols, by exemplifying a linear network with $n=4$. In the $i$-th round ($i =1,2,\ldots,l$), according to the previous outcomes ${\bm k}_{i-1}=k_{i-1} \ldots k_2k_1$, the protocol uses a quantum channel ${\cal N}^{e_{{\bm k}_{i-1}}}$ with $e_{{\bm k}_{i-1}} \in E$, followed by LOCC providing a quantum state $\hat{\rho}_{{\bm k}_i}^{ABC^1C^2\ldots C^n}$ with a new outcome $k_i$. After an $l$-th round, Alice and Bob may obtain the resources for quantum communication.
}
  \label{fig:1}
\end{figure}

In the conventional Internet, 
if a client, Alice, wants to communicate with another client, Bob, an Internet protocol determines the path that the data follows to travel across multiple networks from Alice to Bob. Analogously, in the future, according to a request for performing quantum communication between Alice and Bob, a quantum internet protocol will supply the resources for quantum communication, such as unconditionally secure key and quantum entanglement, to Alice and Bob by utilizing proper intermediate nodes connected by optical fibres with each other \cite{K08} [c.f. Fig.~\ref{fig:1}~(a)].
To such an optical network, photon loss in optical fibres is the dominant impediment in general \cite{L10}.
Nonetheless, as long as Alice and Bob are not too far away from each other, say over a couple hundred kilometres, 
the intermediate nodes would not be necessary, 
because the current point-to-point quantum communication has already been very efficient as well as ready for practical use \cite{review2}.
Besides, in terms of the communication efficiency for the distance,
known schemes \cite{BB84,B92,RRDPS,AK12,A09,H03,W05,LMC05} for the point-to-point links have no scaling gap with a general upper bound on quantum communication capacity and secret key agreement capacity of any memoryless lossy channel (which may or may not be noisy) under the use of unlimited forward and backward classical communication, called Takeoka-Guha-Wilde (TGW) bound \cite{TGW14,TGW14e} (see also subsequent important improvements \cite{PLO15,PL15}  to the TGW bound).
Hence, there remains not much room to improve those schemes for point-to-point links further.
Nevertheless, the point-to-point communication is not efficient enough to achieve the quantum internet.
For example, the point-to-point quantum communication over 1,000~km needs to take almost one century to provide just one bit of secure key or one ebit for them under the use of a typical standard telecom optical fibre with loss of about 0.2~dB/km \cite{G02}.
Therefore, for the request from far distant Alice and Bob, the quantum internet necessitates long-distance quantum communication schemes utilizing intermediate nodes, 
such as intercity QKD protocols \cite{ATM15,AKB14,PRML14} and quantum repeaters \cite{B98,DLCZ,SSRG09,ATKI10,M10,M12,J09,C06,L06,G12,L12,ZDB12,KWD03,MATN15,ATL15}.
In particular, these schemes would be in greater demand for the quantum internet than the point-to-point quantum communication,
analogously to the current Internet where a client communicates with a far distant client via repeater nodes commonly and even unconsciously. 
Therefore, besides the TGW bound for the point-to-point links, it is important to find out a similarly fundamental and general limitation on the long-distance quantum communication schemes, which results in understanding the full potential of the future quantum internet.
However, it remained open to determine this limitation due to the complex nature of those schemes coming from the use of many intermediate nodes as well as a large variety of combinations of different elements such as quantum memories and quantum error correction and of different primitives such as entanglement generation, entanglement swapping and entanglement purification.

The main point of this paper is to present a fundamental and practical limitation on the quantum internet.
In particular, we derive rate-loss tradeoffs for any two-party quantum communication---composed of the use of optical fibres connecting nodes as well as arbitrary local operations and unlimited forward and backward (public) classical communication (LOCC)---over the quantum internet, by tailoring the TGW bound to being applicable to any network topology.
The key insight is reduction. Given any quantum network (which might be a subnetwork of a quantum internet), Alice's node $A$ and Bob's node $B$, we can consider any bipartition of the nodes in the quantum network, $V_A$ including node $A$ and $V_B$ containing node $B$ [c.f.~Fig.~\ref{fig:1} (a)]. By regarding all nodes at $V_A$ as local at $A$ and all nodes at $V_B$ as local at $B$---which could never increase the difficulty of quantum communication between $A$ and $B$, 
one could reduce any network flow as a flow between a point-to-point link between $A$ and $B$ only. Therefore, an upper bound on the key rate or distillable entanglement for a point-to-point link automatically carries over to an upper bound to the quantum network.
As this upper bound for point-to-point links, we simply use the TGW bound \cite{TGW14}. 
Our reduction idea is a simple observation. Nonetheless, rather remarkably, we will show here that the obtained bounds are excellent in the sense that they have no scaling gap with achievable quantum communication efficiencies of known protocols for intercity QKD and quantum repeaters, in terms of rate-loss tradeoffs.
This is brought by the fact that our bounds essentially depend only on the number of the channel uses to establish a quantum communication resource for Alice and Bob and the squashed entanglement \cite{TGW14,TGW14e} of the used optical channels.

To obtain our bound, we need to define a general paradigm of two-party communication over the quantum internet [c.f. Fig.~\ref{fig:1}~(a)].
In the quantum internet, there are a variety of optical channels connecting nodes, for example depending on those lengths.
This necessitates to generalize the paradigm \cite{TGW14,TGW14e} of Takeoka {\it et al.} for the point-to-point communication where it has been enough to treat only one optical channel between Alice and Bob.
For instance,
we need to allow the choice of which channel to use in the next round to depend on the outcomes of LOCC operations in previous rounds, in contrast to the paradigm of Takeoka~{\it et al.} \cite{TGW14,TGW14e}.

To make this more precise, 
let us define the general protocol. 
We assume that any classical communication over the network is freely usable.
Suppose that Alice ($A$) and Bob ($B$) call a quantum internet protocol to share a resource for quantum communication, unconditionally secure key or quantum entanglement, over the quantum network. Accordingly, the quantum internet protocol determines a subnetwork to supply the resource to Alice and Bob. 
The subnetwork is characterized by a directed graph $G=(V,E)$ with a set $V$ of vertices and a set $E$ of edges, where the vertices of $G$ represent Alice's node, Bob's node and intermediate nodes $\{C^k\}_{k=1,2,\ldots,n}$ in the subnetwork, i.e., $V=\{A,B,C^1,C^2,\ldots,C^{n}\}$, and an edge $\varepsilon=v_1\to v_2\in E$ of $G$ for $v_1,v_2 \in V$ specifies a quantum channel ${\cal N}^{v_1\to v_2}$ to send a quantum system from node $v_1$ to node $v_2$ in the subnetwork.
Then, the most general protocol proceeds in an adaptive manner as follows [c.f. Fig.~\ref{fig:1}~(b) which exemplifies a linear network with $n=4$].
The protocol starts by preparing the whole system in a separable state $\hat{\rho}_1^{ABC^1C^2\ldots C^n}$ and then by using a quantum channel ${\cal N}^{e_1}$ with $e_1\in E$. This is followed by arbitrary LOCC among all the nodes, which gives an outcome $k_1$ and a quantum state $\hat{\rho}_{k_1}^{ABC^1C^2\ldots C^n}$ with probability $p_{k_1}$. 
In the second round, depending on the outcome $k_1$, a node uses a quantum channel ${\cal N}^{e_{k_1}}$ with $e_{k_1} \in E$, followed by LOCC among all the nodes. This LOCC gives an outcome $k_2$ and a quantum state $\hat{\rho}_{k_2 k_1}^{ABC^1C^2\ldots C^n}$ with probability $p_{k_2|k_1}$. Similarly, in the $i$-th round, according to the previous outcomes  ${\bm k}_{i-1}:=k_{i-1} \ldots k_2k_1$ (with ${\bm k}_0:=1$), the protocol uses a quantum channel ${\cal N}^{e_{{\bm k}_{i-1}}}$ with $e_{{\bm k}_{i-1}} \in E$, followed by LOCC providing a quantum state $\hat{\rho}_{{\bm k}_i}^{ABC^1C^2\ldots C^n}$ with a new outcome $k_i$ with probability $p_{k_i|{\bm k}_{i-1}}$. After a finite number of rounds, say after an $l$-th round, the protocol must present $\hat{\rho}_{{\bm k}_l}^{AB}={\rm Tr}_{C^1C^2\ldots C^n} ( \hat{\rho}_{{\bm k}_l}^{ABC^1C^2\ldots C^n})$ close to a target state $\hat{\tau}_{d_{{\bm k}_l}}^{AB}$ with rank $d_{{\bm k}_l}$ in the sense of $||\hat{\rho}_{{\bm k}_l}^{AB} -\hat{\tau}_{d_{{\bm k}_l}}^{AB} ||_1 \le \epsilon$ for $\epsilon>0$, 
from which Alice and Bob can distil $\log_2 d_{{\bm k}_l}$ secret bits for the purpose of the unconditionally secure communication or $\log_2 d_{{\bm k}_l}$ ebits for the purpose of the quantum teleportation.
After all, the protocol results in presenting $\log_2 d_{{\bm k}_l}$ secret bits or ebits with probability $p_{{\bm k}_l}$ by using quantum channels $\{ {\cal N}^{e_{{\bm k}_i}} \}_{i=0,1,\ldots,l-1}$,
where $p_{{\bm k}_i}:=p_{k_i| {\bm k}_{i-1}}  \ldots p_{k_3| {\bm k}_2} p_{k_2|k_1} p_{k_1}$.

For this general adaptive protocol, our main result is described as follows.
Let us divide set $V$ into two disjoint sets, $V_A$ including $A$ and $V_B$ including $B$, satisfying $V_A=V \setminus V_B$ and $V_B=V \setminus V_A$ [c.f. Fig.~\ref{fig:1} for the examples].
If ${\cal N}^{e_{{\bm k}_i}}$ is a channel between a node in $V_A$ and a node in $V_B$, we write ${\bm k}_i \in K_{V_A\leftrightarrow V_B}$. For example, ${\bm k}_1\in K_{V_A\leftrightarrow V_B}$ in  Fig.~\ref{fig:1}~(b).
Then, for any choice of $V_A$ and $V_B$, the most general protocol has a limitation described by
\begin{align}
\sum_{{\bm k_l} } p_{{\bm k}_l}  \log_2 d_{{\bm k}_l} \le  \frac{1}{1-16\sqrt{\epsilon}} \left(\sum_{i=0}^{l-1} \sum_{{\bm k_{i}} \in K_{V_A\leftrightarrow V_B}  } p_{{\bm k}_{i}}  E_{\rm sq} ( {\cal N}^{e_{{\bm k}_{i}}})  + 
4 h(2\sqrt{\epsilon}) \right), \label{eq:mc}
\end{align}
where $h$ is the binary entropy function with a property of $\lim_{x \to 0} h(x)=0$ and $E_{\rm sq}({\cal N})$ is the squashed entanglement of channel ${\cal N}$ \cite{TGW14,TGW14e}. 
This bound is reduced to $\sum_{{\bm k_l} } p_{{\bm k}_l}  \log_2 d_{{\bm k}_l} \le \sum_{i=0}^{l-1} \sum_{{\bm k_{i}} \in K_{V_A\leftrightarrow V_B}  } p_{{\bm k}_{i}}  E_{\rm sq} ( {\cal N}^{e_{{\bm k}_{i}}}) $ for $\epsilon \to 0$.
The bound (\ref{eq:mc}) is obtained by regarding the general multi-party protocol as bipartite communication between $V_A$ and $V_B$ and by applying the TGW bound to the bipartite one (see Appendix for the proof).
Since the bound holds for any choice of $V_A$, the bound shows that the average of the obtained secret bits or ebits is most tightly bounded by the choice of $V_A$ that minimizes the right-hand side of Eq.~(\ref{eq:mc}).

As an instructive application of the bound ($\ref{eq:mc}$), we first derive an upper bound for a linear optical network, which includes intercity QKD protocols and quantum repeater protocols.
Here the goal of Alice and Bob, separated over distance $L$, is to share secret bits or ebits by utilizing intermediate nodes $\{C^j\}_{j=1,\ldots,n}$.
For simplicity, suppose that they are located at regular intervals and connected with optical fibres with transmittance $\eta_{L_0}:=e^{-L_0/l_{\rm att}}$ for attenuation length $l_{\rm att}$ and $L_0:=L/(n+1)$ with each other. 
Then, since they can use only the same optical channel connecting the adjacent nodes at best,
all the channels $\{ {\cal N}^{e_{{\bm k}_i}} \}_{i=0,1,\ldots,l-1}$ must be the same lossy channel with transmittance $\eta_{L_0}$, for which Takeoka {\it et al.} have already derived an upper bound on the squashed entanglement of the channel \cite{TGW14,TGW14e}. This implies $E_{\rm sq} ( {\cal N}^{e_{{\bm k}_{i}}}) \le 2 \log_2 [(1+\eta_{L_0})/ (1-\eta_{L_0})]$ for any of the channels $\{ {\cal N}^{e_{{\bm k}_i}} \}_{i=0,1,\ldots,l-1}$, where the factor $2$ in the front comes from the fact that a single use of an optical channel for transmission of an optical pulse corresponds to the sending of two optical modes associated with its polarization degrees of freedom.
Then, the bound (\ref{eq:mc}) is reduced to 
\begin{align}
 \langle \log_2 d_{{\bm k}_l} \rangle_{{\bm k}_l} \le  \frac{1}{1-16\sqrt{\epsilon}} \left[ 2 \log_2 \left( \frac{1+\eta_{L_0}}{1-\eta_{L_0}} \right) \sum_{i=0}^{l-1} \sum_{{\bm k_{i}} \in K_{V_A\leftrightarrow V_B}  } p_{{\bm k}_{i}}   + 
4 h(2\sqrt{\epsilon}) \right], \label{eq:qr-1}
\end{align}
where $\langle f_{{\bm k}_l} \rangle_{{\bm k}_l}$ represents the average of function $f_{{\bm k}_l} $ over ${\bm k}_l$, that is, $\langle f_{{\bm k}_l} \rangle_{{\bm k}_l}:= \sum_{{\bm k_l} } p_{{\bm k}_l}  f_{{\bm k}_l}  $.
For the choice of $V_A=\{A\}$ and $V_B=\{C^1,C^2,\ldots,C^n, B\}$ ($V_A=\{A,C^1,C^2,\ldots,C^n\}$ and $V_B=\{B\}$), $\sum_{i=0}^{l-1} \sum_{{\bm k_{i}} \in K_{V_A\leftrightarrow V_B}  } p_{{\bm k}_{i}}$ represents the average number $\langle m^{A \leftrightarrow C^1}_{{\bm k}_l} \rangle_{{\bm k}_l}$ ($\langle m^{C^n \leftrightarrow B}_{{\bm k}_l} \rangle_{{\bm k}_l}$) of times the optical channel between Alice and node $C^1$ (between node $C^n$ and Bob) is used.
Similarly, for the choice of  $V_A=\{A,C^1,\ldots,C^{j}\}$ and $V_B=\{C^{j+1},\ldots,C^n, B\}$ for $j=1,2,\ldots, n-1$,  $\sum_{i=0}^{l-1} \sum_{{\bm k_{i}} \in K_{V_A\leftrightarrow V_B}  } p_{{\bm k}_{i}}$ describes the average number $\langle m_{{\bm k}_l}^{C^{j} \leftrightarrow C^{j+1}} \rangle_{{\bm k}_l}$ of channel uses between node $C^{j}$ and node $C^{j+1}$.
Hence, Eq.~(\ref{eq:qr-1}) gives
\begin{align}
 \langle \log_2 d_{{\bm k}_l} \rangle_{{\bm k}_l} \le  \frac{1}{1-16\sqrt{\epsilon}} \left[ 2 \min\{\langle  m_{{\bm k}_l}^{A \leftrightarrow C^1} \rangle_{{\bm k}_l}, 
\langle  m_{{\bm k}_l}^{C^1 \leftrightarrow C^2} \rangle_{{\bm k}_l}, \ldots,
\langle  m_{{\bm k}_l}^{C^n \leftrightarrow B} \rangle_{{\bm k}_l}\}  \log_2 \left( \frac{1+\eta_{L_0}}{1-\eta_{L_0}} \right)  + 
4 h(2\sqrt{\epsilon}) \right]. \label{eq:qr-2}
\end{align}
Since $\langle  m_{{\bm k}_l}^{A \leftrightarrow C^1} \rangle_{{\bm k}_l}+ 
\langle  m_{{\bm k}_l}^{C^1 \leftrightarrow C^2} \rangle_{{\bm k}_l}+\ldots + 
\langle  m_{{\bm k}_l}^{C^n \leftrightarrow B} \rangle_{{\bm k}_l}
=l$, we have $\min\{\langle  m_{{\bm k}_l}^{A \leftrightarrow C^1} \rangle_{{\bm k}_l}, 
\langle  m_{{\bm k}_l}^{C^1 \leftrightarrow C^2} \rangle_{{\bm k}_l}, \ldots,
\langle  m_{{\bm k}_l}^{C^n \leftrightarrow B} \rangle_{{\bm k}_l}\} \le l/(n+1)$,
which concludes 
\begin{align}
 \langle \log_2 d_{{\bm k}_l} \rangle_{{\bm k}_l} \le  \frac{1}{1-16\sqrt{\epsilon}} \left[ \frac{2l}{n+1}  \log_2 \left( \frac{1+\eta_{L_0}}{1-\eta_{L_0}} \right) + 
4 h(2\sqrt{\epsilon}) \right]. \label{eq:qr-3}
\end{align}
In particular, this bound shows that the average secret bits or ebits per channel use, $ \langle \log_2 d_{{\bm k}_l} \rangle_{{\bm k}_l} /l$, are upper bounded by $2(n+1)^{-1}\log_2[(1+\eta_{L_0})/(1-\eta_{L_0})] $ for $\epsilon \simeq 0$, which is approximated to be $4[ (n+1)\ln 2]^{-1}  \eta_{L_0}$ for $L_0 \gg 1$.
The bound (\ref{eq:qr-3}) is tight enough to show that the existing intercity QKD protocols and quantum repeater protocols are pretty good in the sense that they have the same scaling with this simple bound.

\begin{figure}
\includegraphics[keepaspectratio=true,height=45mm]{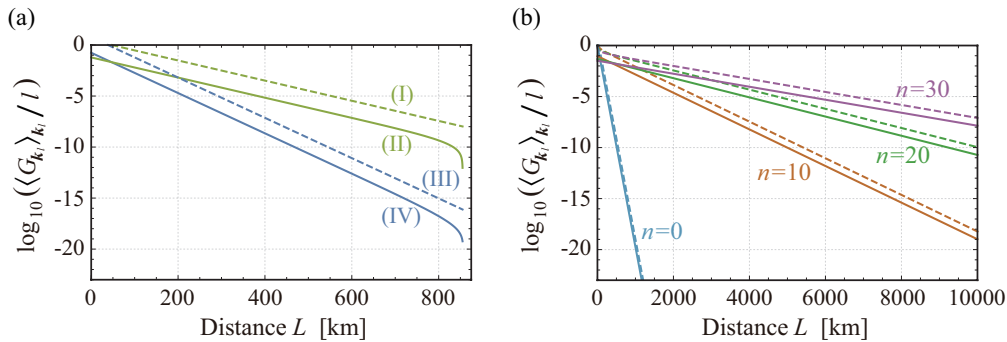}
  \caption{Secret bits or ebits per channel use, $\langle G_{{\bm k}_l} \rangle_{{\bm k}_l} /l$ with $G_{{\bm k}_l}:=\log_{2} d_{{\bm k }_l}$, for protocols based on a linear network with total distance $L$. The protocols use intermediate nodes $\{C^1,C^2,\ldots,C^n\}$ connected by optical fibres each other and located at regular intervals, say $L_0=L/(n+1)$.
The solid curves represent achievable performance, while the dashed curves are our general upper bounds (\ref{eq:qr-3}) for the linear network, $2(n+1)^{-1}\log_2[(1+\eta_{L_0})/(1-\eta_{L_0})] $, for various $n$.
In panel (a), we provide the performance of mdiQKD protocols \cite{ATM15,LCQ12} using only a single intermediate node ($n=1$) equipped with feasible optical devices.
Lines (II) and (IV) represent the all-photonic intercity QKD protocol \cite{ATM15} and the original mdiQKD protocol \cite{LCQ12}, respectively. These lines just refer to the performance given in Fig.~3 of Ref.~\cite{ATM15}, where a collection of the state-of-art optical devices such as single-photon sources, active feedforward technique with an optical switch and single-photon detectors is assumed to be used (c.f. \cite{ATM15} for the details). 
The key rate scales linearly with $\eta_L$ (i.e., the
transmittance over distance $L$) for the mdiQKD \cite{LCQ12}, but
it scales linearly with the square root of $\eta_L$ for the all-photonic intercity QKD \cite{ATM15}.
In addition, we show our general bound (\ref{eq:qr-3}) for $n=1$ as line (I) and the TGW bound \cite{TGW14} (corresponding to our bound with $n=0$) as line (III). 
Comparing lines (I) and (II), we can see that the all-photonic intercity QKD protocol has the same scaling with our general bound (\ref{eq:qr-3}) for $n=1$. 
In panel (b), for various $n$, we provide the performance of the idealized qubit-based quantum repeater protocol introduced in the main text, $(n+1)^{-1} \eta_{L_0}=(n+1)^{-1} \eta_{L/(n+1)}$, as solid lines and our bound (\ref{eq:qr-3}) as dashed curves. We can see that there is essentially no scaling gap between our bound (\ref{eq:qr-3}) and the idealized qubit-based protocol. 
}
  \label{fig:2}
\end{figure}

To show this, let us first compare the bound (\ref{eq:qr-3}) with the intercity QKD protocols \cite{ATM15,AKB14,PRML14}.  
This class of QKD protocols leads to a square root improvement in the secret key rate over conventional QKD schemes (without quantum repeaters) bounded by the TGW bound.
Nonetheless, it can be obtained without the need of matter quantum memories or quantum error correction \cite{ATM15}, which is in a striking contrast to quantum repeaters \cite{B98,DLCZ,SSRG09,ATKI10,M10,M12,J09,C06,L06,G12,L12,ZDB12,KWD03,MATN15,ATL15}.
In particular, 
those protocols are modifications of the measurement-device-independent QKD (mdiQKD) \cite{LCQ12} and all of them use a single untrusted intermediate node $C$ in the middle of communicators Alice and Bob.
Node $C$ shares optical channels with Alice and Bob, whose transmittance is described by $\eta_{L/2}$.
Then, using matter quantum memories at node $C$ \cite{AKB14,PRML14} or using only optical devices at node $C$ \cite{ATM15}, 
the protocols present a secret bit per about $\eta_{L/2}^{-1}$ uses of optical channels between Alice and node $C$ and between Bob and node $C$.
This implies that the average secret bits of these protocols per channel use are in the order of $\eta_{L/2}$.
However, this is exactly the same scaling of the bound (\ref{eq:qr-3}), because the bound (\ref{eq:qr-3}) is proportional to $\eta_{L_0}=\eta_{L/2}$ for $n=1$, $\epsilon \simeq 0 $ and $L_0 \gg 1$.
In fact, this is easily confirmed by seeing Fig.~\ref{fig:2}~(a). 

Next, let us compare the bound (\ref{eq:qr-3}) with the performance of achievable quantum repeater protocols. Actually, there are many quantum repeater schemes \cite{B98,DLCZ,SSRG09,ATKI10,M10,M12,J09,C06,L06,G12,L12,ZDB12,KWD03,MATN15,ATL15}, depending on the assumed devices of the repeater nodes $\{ C^1,C^2,\ldots,C^{n}\}$.
For instance, a protocol assumes repeater nodes equipped with atomic-ensemble quantum memories as well as optical devices \cite{DLCZ,SSRG09}. 
To obtain better scaling,
instead of the atomic-ensemble quantum memories,
another protocols \cite{B98,M10,M12,J09,G12} use matter qubits satisfying not only DiVincenzo's five criteria for a universal quantum computation but also his extra criterion on the efficient coupling with single photons \cite{D00}. Moreover, recently, there is even an all-photonic scheme \cite{ATL15} that does not use matter quantum memories at all and works by using only optical devices. 
However, since our aim here is to show the existence of a quantum repeater protocol that has the same scaling with the bound (\ref{eq:qr-3}) in principle,
let us introduce an idealized qubit-based protocol which uses a noiseless quantum computer with the function of the perfect coupling with single photons at each repeater node.

In the idealized qubit-based protocol, (i) a node $X \in V$ except for $B$ begins by producing a single photon which is in maximally entangled state $\ket{\Phi^+}=(\ket{0}\ket{H} + \ket{1}\ket{V})/\sqrt{2}$ with a qubit of the local quantum computer, where $\{\ket{H},\ket{V}\}$ is an orthonormal basis for the polarization degrees of freedom of the single photon and $\{\ket{0},\ket{1}\}$ is a computational basis of the qubit.
(ii) Then, the node $X$ except for $B$ sends its right-hand-side adjacent node the single photon through the optical fibre with transmittance $\eta_{L_0}$. 
(iii) On receiving the photon from the left-hand-side adjacent node, the node $X$ except for $A$ performs a quantum non-demolition (QND) measurement to confirm the successful arrival of the single photon, and announces the measurement outcome via a heralding signal.
If this QND measurement proves the successful arrival of the single photon, the node $X$ transfers the quantum state of the photon into a qubit of the local quantum computer faithfully, establishing a maximally entangled state between the node $X$ and the left-hand-side adjacent node.
(iv) If the node $X$ except for $B$ is informed of the loss of the sent photon in the transmission by the heralding signal from the right-hand-side adjacent node, the node $X$ and its right-hand-side adjacent node repeat steps (i)-(iii). 
(v) If every node shares a maximally entangled state with the adjacent nodes, all the repeater nodes $\{C^1,C^2,\ldots,C^{n} \}$ apply the Bell measurement to a pair of local qubits that have been entangled with qubits in the adjacent repeater nodes. This gives Alice and Bob a pair of qubits in a maximally entangled state.

Let us estimate the performance of this idealized qubit-based protocol.
Since the entanglement generation process (i)-(iii) is repeated until a single photon sent in step (ii) survives over the fibre transmission with transmittance $\eta_{L_0}$,
the average of the number $m$ of the channel uses to obtain the entanglement between adjacent nodes in step (iii) is $\sum_{m=1}^{\infty} m (1-\eta_{L_0})^{m-1} \eta_{L_0}=\eta_{L_0}^{-1}$.
If we regard the entanglement generation process (i)-(iii) as a single round of the protocol and the process is executed between adjacent repeater nodes in parallel independently, 
the idealized qubit-based protocol should present a pair of qubits in a maximally entangled state for the total number $l$ of the rounds with $l = (n+1) \eta_{L_0}^{-1}$ in a manner arbitrary close to a deterministic process.
Therefore, the average secret bits or ebits of the idealized qubit-based protocol per channel use is $(n+1)^{-1} \eta_{L_0}$, 
which is exactly the same scaling of the bound (\ref{eq:qr-3}).
This fact is also easily confirmed by seeing Fig.~\ref{fig:2}~(b). 

Since the existing quantum repeater protocols \cite{B98,DLCZ,SSRG09,ATKI10,M10,M12,J09,C06,L06,G12,L12,ZDB12,KWD03,MATN15,ATL15} are based on more practical devices than the idealized qubit-based protocol,
they must be less efficient than the idealized qubit-based protocol, owing to more imperfections caused by the practical devices.
However, there are schemes \cite{M12,ATL15,M10,J09,G12} whose performance is essentially determined by distance $L_0$ even under the use of such more practical devices similarly to the idealized qubit-based protocol as well as our bound (\ref{eq:qr-3}). 
In other words, the quantum repeater protocols \cite{M10,J09,G12,M12,ATL15} have no scaling gap with our bound (\ref{eq:qr-3}).

We have seen that our bound (\ref{eq:mc}) is simple but powerful enough to derive rate-loss tradeoffs (\ref{eq:qr-3}) with the same scaling with existing intercity QKD protocols and quantum repeaters. 
Note that we have assumed that the intermediate nodes $\{C^1,C^2,\ldots,C^{n} \}$ are located at regular intervals, but this assumption is unnecessary.
In particular, similarly to the derivation of Eq.~(\ref{eq:qr-3}),
from our bound (\ref{eq:mc}), it is not difficult to obtain rate-loss tradeoffs for any linear optical network where the intermediate nodes are not necessarily positioned at regular intervals, and the existing intercity QKD protocols and quantum repeater protocols are still shown to have the same scaling with the tradeoffs.
Moreover, if the future quantum internet protocol was programmed to always find out a linear network as the subnetwork $G$ over the quantum internet by using an algorithm for the shortest path problem, our bound for the linear network would help to run the algorithm via determining the weights for all the edges associated with optical channels.
This application will also hold even if the quantum internet protocol uses tree networks as the subnetwork $G$ instead of linear networks.
More generally, the quantum internet protocol prefers to use a more general network topology as the subnetwork $G$. Even in this case, our bound is useful to determine the obtainable secret bits or ebits because they are always upper bounded by Eq.~(\ref{eq:mc}) for the minimum bipartition given by choosing $V_A$ and $V_B$ properly.

Although we have just considered only the lossy channels as the examples of the applications of our bound (\ref{eq:mc}), 
this is, of course, not only the case as long as the squashed entanglement of the channel in Eq.~(\ref{eq:mc}) can be estimated for given channels. 
Even though in Fig.~\ref{fig:2}, we have plotted only the results for a linear chain of pure-loss channels, it should be noted that the TGW bound is very general and applies to any memoryless lossy channels (which may or may not be noisy) and, with the observation made in the present paper, to any network topology.
While we have employed mainly the TGW bound in our paper, it should be noted that our reduction idea may be useful to derive a good bound for a general network topology from a bound for point-to-point quantum communication generally.
We have just begun to grasp full implications of our bound: For instance, its applications to the many-body physics regarded as a quantum network and to a more complicated quantum communication protocol---such as a multi-party protocol like Ref.~\cite{STW15}---will be in a fair way to appear.

We thank S.~Guha, S. Pirandola, M.~Takeoka and M.~M.~Wilde for valuable discussions about their papers \cite{TGW14, TGW14e, PL15,PLO15,P16}.
K.A. thanks support from the Project UQCC by the National Institute of
Information and Communications Technology.
H.-K.L. acknowledges financial support from
NSERC and CRC program.

{\it Note added.}---During the preparation of this paper, Pirandola uploaded a related paper \cite{P16} on the arXiv, based on their recent papers \cite{PL15,PLO15}.
Our results do not subsume, nor are they subsumed by the results in Ref.~\cite{P16}. Pirandola assumed that the channels
are {\it stretchable} whereas we do not make such an assumption \cite{note1}.
On the other hand, for the specific case of a purely lossy channel that is used in generating the simulation results in Fig.~\ref{fig:2},
Pirandola's result gives a better bound than ours.

\appendix
\section{Proof for the main result (1)}

Here we provide the proof for the main result, that is, Eq.~(1).
Although here we focus on deriving the bound (1) for QKD protocols between Alice and Bob, 
the same technique can be applied for protocols to share entanglement between them, similarly to the TGW bound \cite{TGW14,TGW14e}.

Suppose that, with the help of other parties $\{C^k \}_{k=1,2,\ldots,n}$ in a quantum network, Alice and Bob share physical systems ${\cal H}_{A'A''} \otimes {\cal H}_{B'B''}$ in private state \cite{HHHO05}
\begin{equation}
\hat{\gamma}^{AB}_d=\hat{U}^{A'B'A''B''} (\ket{\Phi} \bra{\Phi}_{A'B'} \otimes \hat{\rho}^{A''B''}) \hat{U}^{A'B'A''B''\dag}
\end{equation}
with unitary operator $\hat{U}^{A'B'A''B''}:=\sum_{i,j=0}^{d-1} \ket{ij} \bra{ij}_{A'B'} \otimes \hat{U}^{A''B''}_{ij}$, maximally entangled state $\ket{\Phi}_{A'B'}:= \sum_{i=0}^{d-1} \ket{ii}_{A'B'}/\sqrt{d}$ and orthonormal states $\{\ket{ij}_{A'B'}\}_{i,j=0,1,\ldots,d-1}$ for systems ${\cal H}_{A'} \otimes {\cal H}_{B'}$.
In particular, Alice and Bob obtain a private state through the following most general adaptive protocol:
(i) Alice, Bob and parties $\{C^k \}_{k=1,2,\ldots,n}$ begin by preparing their physical systems ${\cal H}^0$ in a separable state $\hat{\rho}_1^{ABC^1C^2\ldots C^n}$, where ${\cal H}^j:={\cal H}_{A}^{j} \otimes {\cal H}_{B}^j \otimes {\cal H}_{C^1}^j \otimes {\cal H}_{C^2}^j\otimes \cdots \otimes {\cal H}_{C^n}^j$  
and 
${\cal H}_{X}^j$ represents the physical system held by party $X\in \{ A,B,C^1,C^2,\ldots, C^n\}$.
(ii) In the first round, party $X|1 \in \{A,B,C^1,C^2,\ldots, C^n \}$ sends his/her subsystem $\bar{{\cal H}}_{X|1}$ to party $Y|1$ through quantum channel ${\cal N}^{\bar{{\cal H}}_{X|1} \to \tilde{{\cal H}}_{Y|1}} $ with isometric extension ${\cal U}^{\bar{{\cal H}}_{X|1}\to  \tilde{{\cal H}}_{Y|1} \otimes {\cal H}_{E_1}}$ for environment system ${\cal H}_{E_1}$, which provides a refreshed description of the whole system, ${\cal H}^{0'|1}$ with ${\cal H}_{Y|1}^{0'|1}={\cal H}_{Y|1}^0 \otimes \tilde{{\cal H}}_{Y|1}$, ${\cal H}_{X|1}^{0}={\cal H}_{X|1}^{0'|1} \otimes \bar{{\cal H}}_{X|1}$ and ${\cal H}_{Z}^{0'|1}={\cal H}_{Z}^0$ for any party $Z$ except for parties $X|1$ and $Y|1$.
This is followed by an LOCC operation, which presents a renewed entire system ${\cal H}^{1|1}$ in state $\hat{\rho}_{k_1}^{ABC^1C^2\ldots C^n}$ with probability $p_{k_1}$. 
Let ${\cal H}_{R_{k_1}}$ be a system that purifies the state $\hat{\rho}_{k_1}^{ABC^1C^2\ldots C^n}$, providing pure-state expression $\ket{\hat{\rho}_{k_1}}_{ABC^1C^2\ldots C^nR_{k_1}}$.
(iii) Similarly, in the $i$th round ($i=2,3,\ldots,l$), 
depending on the previous outcomes ${\bm k}_{i-1}:={ k}_{i-1}\cdots { k}_1$ (with ${\bm k}_{0}:=1$), for given entire system ${\cal H}^{(i-1)|{\bm k}_{i-2}}$,
party $X| {\bm k}_{i-1} \in \{A,B,C^1,C^2,\ldots, C^n \}$ sends his/her subsystem $\bar{{\cal H}}_{X|{{\bm k}_{i-1}}}$ to party $Y|{{\bm k}_{i-1}}\in \{A,B,C^1,C^2,\ldots, C^n \}$ through quantum channel ${\cal N}^{\bar{{\cal H}}_{X|{{\bm k}_{i-1}}} \to \tilde{{\cal H}}_{Y|{{\bm k}_{i-1}}} } $ with isometric extension ${\cal U}^{\bar{{\cal H}}_{X|{{\bm k}_{i-1}}} \to \tilde{{\cal H}}_{Y|{{\bm k}_{i-1}}}\otimes {\cal H}_{E_{{\bm k}_{i-1}}}}$ for environment system $ {\cal H}_{E_{{\bm k}_{i-1}}}$, which updates the description of the whole system as ${\cal H}^{(i-1)'|{\bm k}_{i-1}}$ with ${\cal H}^{(i-1)'|{\bm k}_{i-1}}_{Y|{{\bm k}_{i-1}}}={\cal H}^{(i-1)|{\bm k}_{i-2}}_{Y|{{\bm k}_{i-1}}} \otimes \tilde{{\cal H}}_{Y|{{\bm k}_{i-1}}}$, ${\cal H}^{(i-1)|{{\bm k}_{i-2}}}_{X|{{\bm k}_{i-1}}}={\cal H}^{(i-1)'|{{\bm k}_{i-1}} }_{X|{{\bm k}_{i-1}}} \otimes\bar{ {\cal H}}_{X|{{\bm k}_{i-1}}}$ and ${\cal H}_{Z}^{(i-1)'|{\bm k}_{i-1}}={\cal H}_{Z}^{(i-1)|{{\bm k}_{i-2}}}$ for any party $Z$ except for parties $X|{{\bm k}_{i-1}}$ and $Y|{{\bm k}_{i-1}}$.
This is followed by an LOCC operation, providing an entire system ${\cal H}^{i|{\bm k}_{i-1}}$ in state $\hat{\rho}_{{\bm k}_i}^{ABC^1C^2\ldots C^n}$ with probability $p_{k_i| {\bm k}_{i-1}}$. Let ${\cal H}_{R_{{\bm k}_{i}}}$ be a system that purifies the state $\hat{\rho}_{{\bm k}_i}^{ABC^1C^2\ldots C^n}$, presenting pure-state expression $\ket{\hat{\rho}_{{\bm k}_i}}_{ABC^1C^2\ldots C^n R_{{\bm k}_{i}} }$.
(iv) Finally, i.e., in the $l$th round, Alice and Bob obtain state $\hat{\rho}_{{\bm k}_l}^{ABC^1C^2\ldots C^n}$ close to private state $\hat{\gamma}_{d_{{\bm k}_l}}^{AB}$ for integer $d_{{\bm k}_l}(\ge 1)$.

From the definition, the final state $\hat{\rho}_{{\bm k}_l}^{ABC^1C^2\ldots C^n}$ should be close to private state $\hat{\gamma}^{AB}_{d_{{\bm k}_l}}$, i.e., $||\hat{\rho}_{{\bm k}_l}^{AB}- \hat{\gamma}^{AB}_{d_{{\bm k}_l}}||_1 \le \epsilon$ for $\epsilon >0$, where we define $\hat{\rho}^X:={\rm Tr}_Y (\hat{\rho}^{XY})$.
From the continuity of the squashed entanglement \cite{C06phd}, this implies 
\begin{equation}
|E_{\rm sq}^{{\cal H}_A^{l|{\bm k}_{l-1}}:{\cal H}_B^{l|{\bm k}_{l-1}}} (\hat{\rho}_{{\bm k}_l}^{AB})- E_{\rm sq}^{{\cal H}_A^{l|{\bm k}_{l-1}}:{\cal H}_B^{l|{\bm k}_{l-1}}} (\hat{\gamma}^{AB}_{d_{{\bm k}_l}})| \le 16\sqrt{\epsilon} \log d'_{{\bm k}_{l}} +4 h(2\sqrt{\epsilon}),
\end{equation}
where $d'_{{\bm  k}_l}:=\min\{ \dim( {\cal H}_A^{l|{\bm k}_{l-1}} ),\dim( {\cal H}_B^{l|{\bm k}_{l-1}}) \}$, $h(x):=-x \log_2 x -(1-x)\log_2 (1-x)$ and $E_{\rm sq}^{X:Y} (\hat{\rho}^{XY})$ is the squashed entanglement between systems $X$ and $Y$ in state $\hat{\rho}^{XY}$ \cite{CW04}.
Since $d'_{{\bm k}_l}=d_{{\bm k}_l}$ without loss of generality and $ E_{\rm sq}^{{\cal H}_A^{l|{\bm k}_{l-1}}:{\cal H}_B^{l|{\bm k}_{l-1}}} (\hat{\gamma}^{AB}_{d_{{\bm k}_l}})\ge \log d_{{\bm k}_l}$ \cite{C06phd}, we have
\begin{align}
\log d_{{\bm k}_l} \le  \frac{1}{1-16\sqrt{\epsilon}} (E_{\rm sq}^{{\cal H}_A^{l|{\bm k}_{l-1}}:{\cal H}_B^{l|{\bm k}_{l-1}}} (\hat{\rho}_{{\bm k}_l}^{AB})+ 4 h(2\sqrt{\epsilon})). \label{eq:proof1}
\end{align}

Our proof for Eq.~(1) is made by regarding the general multi-party protocol as bipartite communication and by applying the technique of the TGW bound \cite{TGW14} to the bipartite one.
Hence, let us divide the set of parties $\{ A,B,C^1,C^2,\ldots, C^n\} (=:{\cal P})$ into two disjoint groups ${\cal P}_A$ and ${\cal P}_B(={\cal P}\setminus {\cal P}_A)$ that include parties $A$ and $B$, respectively.
We define ${\cal H}_{{\cal P}_A}^r:= \otimes_{X \in {\cal P}_A} {\cal H}_X^r$ and ${\cal H}_{{\cal P}_B}^r:= \otimes_{X \in {\cal P}_B} {\cal H}_X^r$.
In addition, we regard ${\bm k}_{i-1}$ as ${\bm k}_{i-1} \in K_{\rm in|{\cal P}_A} $ (${\bm k}_{i-1} \in K_{{\rm out}|{\cal P}_A} $) if $X|{{\bm k}_{i-1}} \in {\cal P}_C$ and $Y|{{\bm k}_{i-1}} \in {\cal P}_C$ (if $X|{{\bm k}_{i-1}} \in {\cal P}_C$ and $Y|{{\bm k}_{i-1}} \in {\cal P} \setminus {\cal P}_C$) for $C=A$ or $C=B$.
In what follows, we derive inequalities for two cases, ${\bm k}_{i-1} \in K_{\rm in|{\cal P}_A} $ and ${\bm k}_{i-1} \in K_{{\rm out}|{\cal P}_A} $.

Let us consider an $i$th round with ${\bm k}_{i-1} \in K_{\rm in|{\cal P}_A} $. 
In this case, the channel $ {\cal N}^{\bar{{\cal H}}_{X|{{\bm k}_{i-1}}} \to \tilde{{\cal H}}_{Y|{{\bm k}_{i-1}}}}$ should be regarded as just a local channel for the bipartite communication between ${\cal P}_A$ and ${\cal P}_B$.
To make this clearer, let us first assume $X|{{\bm k}_{i-1}} \in {\cal P}_A$ and $Y|{{\bm k}_{i-1}} \in {\cal P}_A$. Then, we have
\begin{align}
\sum_{k_i} p_{k_i|{\bm k}_{i-1}} E_{\rm sq}^{{\cal H}_A^{i|{{\bm k}_{i-1}}}:{\cal H}_B^{i|{{\bm k}_{i-1}}}} (\hat{\rho}_{{\bm k}_i}^{AB})\le & \sum_{k_i} p_{k_i|{\bm k}_{i-1}} E_{\rm sq}^{{\cal H}_{{\cal P}_A}^{i|{{\bm k}_{i-1}}}:{\cal H}_{{\cal P}_B}^{i|{{\bm k}_{i-1}}}} (\hat{\rho}_{{\bm k}_i}^{ABC^1C^2\ldots C^n}) \\
\le &  E_{\rm sq}^{{\cal H}_{{\cal P}_A}^{(i-1)'|{{\bm k}_{i-1}}}:{\cal H}_{{\cal P}_B}^{(i-1)'|{{\bm k}_{i-1}}}} ( {\cal N}^{\bar{{\cal H}}_{X|{{\bm k}_{i-1}}} \to \tilde{{\cal H}}_{Y|{{\bm k}_{i-1}}}}(\hat{\rho}_{{\bm k}_{i-1}}^{ABC^1C^2\ldots C^n})) \\
\le &  E_{\rm sq}^{{\cal H}_{{\cal P}_A}^{(i-1)|{{\bm k}_{i-2}}}:{\cal H}_{{\cal P}_B}^{(i-1)|{{\bm k}_{i-2}}}} (\hat{\rho}_{{\bm k}_{i-1}}^{ABC^1C^2\ldots C^n}). \label{eq:a1}
\end{align}
The first inequality is derived from the fact that the squashed entanglement does not increase under partial traces.
The second inequality comes from the fact that the squashed entanglement does not increase on average under LOCC.
The final inequality states that the squashed entanglement does not increase under any local quantum channel.
The same inequality is obtained if we begin by assuming $X|{{\bm k}_{i-1}} \in {\cal P}_B$ and $Y|{{\bm k}_{i-1}} \in {\cal P}_B$.

Let us consider an $i$th round with ${\bm k}_{i-1} \in K_{\rm out|{\cal P}_A} $.
In this case, $ {\cal N}^{\bar{{\cal H}}_{X|{{\bm k}_{i-1}}} \to \tilde{{\cal H}}_{Y|{{\bm k}_{i-1}}}}$ is a channel connecting parties ${\cal P}_A$ and ${\cal P}_B$ nontrivially,
which should put a limitation on the communication.
To make this more precise, we first assume $X|{{\bm k}_{i-1}} \in {\cal P}_A$ and $Y|{{\bm k}_{i-1}} \in {\cal P}_B$. Then, we have
\begin{align}
\sum_{k_i}&\; p_{k_i|{\bm k}_{i-1}} E_{\rm sq}^{{\cal H}_A^{i|{{\bm k}_{i-1}}}:{\cal H}_B^{i|{{\bm k}_{i-1}}}} (\hat{\rho}_{{\bm k}_i}^{AB})\le  \sum_{k_i} p_{k_i|{\bm k}_{i-1}} E_{\rm sq}^{{\cal H}_{{\cal P}_A}^{i|{{\bm k}_{i-1}}}:{\cal H}_{{\cal P}_B}^{i|{{\bm k}_{i-1}}}} (\hat{\rho}_{{\bm k}_i}^{ABC^1C^2\ldots C^n}) \\
\le&  E_{\rm sq}^{{\cal H}_{{\cal P}_A}^{(i-1)'|{{\bm k}_{i-1}}}:{\cal H}_{{\cal P}_B}^{(i-1)'|{{\bm k}_{i-1}}}} ( {\cal N}^{\bar{{\cal H}}_{X|{{\bm k}_{i-1}}} \to \tilde{{\cal H}}_{Y|{{\bm k}_{i-1}}}}(\hat{\rho}_{{\bm k}_{i-1}}^{ABC^1C^2\ldots C^n}))  \\
=&   E_{\rm sq}^{{\cal H}_{{\cal P}_A }^{(i-1)'|{{\bm k}_{i-1}}}:{\cal H}_{{\cal P}_B\setminus (Y|{\bm k}_{i-1})}^{(i-1)'|{{\bm k}_{i-1}}} \otimes {\cal H}_{Y|{\bm k}_{i-1}}^{(i-1)|{{\bm k}_{i-2}}}
\otimes \tilde{{\cal H}}_{Y|{\bm k}_{i-1}} } ( {\cal U}^{\bar{{\cal H}}_{X|{{\bm k}_{i-1}}} \to \tilde{{\cal H}}_{Y|{{\bm k}_{i-1}}} \otimes {\cal H}_{E_{{\bm k}_{i-1}}}}(\ket{ \hat{\rho}_{{\bm k}_{i-1}}}_{ABC^1C^2\ldots C^n R_{{\bm k}_{i-1}}} )) \\
\le& E_{\rm sq}^{{\cal H}_{{\cal P}_A }^{(i-1)'|{{\bm k}_{i-1}}}  \otimes {\cal H}_{{\cal P}_B\setminus (Y|{\bm k}_{i-1})}^{(i-1)'|{{\bm k}_{i-1}}} \otimes {\cal H}_{Y|{\bm k}_{i-1}}^{(i-1)|{{\bm k}_{i-2}}} \otimes {\cal H}_{R_{{\bm k}_{i-1}}}
: \tilde{{\cal H}}_{Y|{\bm k}_{i-1}} } ( {\cal U}^{\bar{{\cal H}}_{X|{{\bm k}_{i-1}}} \to \tilde{{\cal H}}_{Y|{{\bm k}_{i-1}}} \otimes {\cal H}_{E_{{\bm k}_{i-1}}}}(\ket{ \hat{\rho}_{{\bm k}_{i-1}}}_{ABC^1C^2\ldots C^n R_{{\bm k}_{i-1}}} )) \nonumber \\
&+ E_{\rm sq}^{{\cal H}_{{\cal P}_A }^{(i-1)'|{{\bm k}_{i-1}}} \otimes \tilde{{\cal H}}_{Y|{\bm k}_{i-1}}  \otimes {\cal H}_{E_{{\bm k}_{i-1}}} :{\cal H}_{{\cal P}_B\setminus (Y|{\bm k}_{i-1})}^{(i-1)'|{{\bm k}_{i-1}}} \otimes {\cal H}_{Y|{\bm k}_{i-1}}^{(i-1)|{{\bm k}_{i-2}}} }  ( {\cal U}^{\bar{{\cal H}}_{X|{{\bm k}_{i-1}}} \to \tilde{{\cal H}}_{Y|{{\bm k}_{i-1}}} \otimes {\cal H}_{E_{{\bm k}_{i-1}}}}(\ket{ \hat{\rho}_{{\bm k}_{i-1}}}_{ABC^1C^2\ldots C^n R_{{\bm k}_{i-1}}} )) \\
=&E_{\rm sq}^{{\cal H}_{{\cal P}_A }^{(i-1)'|{{\bm k}_{i-1}}}  \otimes {\cal H}_{{\cal P}_B\setminus (Y|{\bm k}_{i-1})}^{(i-1)|{{\bm k}_{i-2}}} \otimes {\cal H}_{Y|{\bm k}_{i-1}}^{(i-1)|{{\bm k}_{i-2}}} \otimes {\cal H}_{R_{{\bm k}_{i-1}}}
: \tilde{{\cal H}}_{Y|{\bm k}_{i-1}} } ( {\cal N}^{\bar{{\cal H}}_{X|{{\bm k}_{i-1}}} \to \tilde{{\cal H}}_{Y|{{\bm k}_{i-1}}} }(\ket{ \hat{\rho}_{{\bm k}_{i-1}}}_{ABC^1C^2\ldots C^n R_{{\bm k}_{i-1}}} )) \nonumber \\
&+ E_{\rm sq}^{{\cal H}_{{\cal P}_A }^{(i-1)'|{{\bm k}_{i-1}}} \otimes \bar{{\cal H}}_{X|{{\bm k}_{i-1}}} :{\cal H}_{{\cal P}_B\setminus (Y|{\bm k}_{i-1})}^{(i-1)|{{\bm k}_{i-2}}} \otimes {\cal H}_{Y|{\bm k}_{i-1}}^{(i-1)|{{\bm k}_{i-2}}} }  (\ket{ \hat{\rho}_{{\bm k}_{i-1}}}_{ABC^1C^2\ldots C^n R_{{\bm k}_{i-1}}} ) \\
\le& E_{\rm sq} ( {\cal N}^{\bar{{\cal H}}_{X|{{\bm k}_{i-1}}} \to \tilde{{\cal H}}_{Y|{{\bm k}_{i-1}}} }) 
+E_{\rm sq}^{{\cal H}_{{\cal P}_A }^{(i-1)|{{\bm k}_{i-2}}}  :{\cal H}_{{\cal P}_B}^{(i-1)|{{\bm k}_{i-2}}} }  (\hat{\rho}_{{\bm k}_{i-1}}^{ABC^1C^2\ldots C^n} ) . \label{eq:a2}
\end{align}
The first inequality is derived from the fact that the squashed entanglement does not increase under partial traces.
The second inequality comes from the fact that the squashed entanglement does not decrease on average under LOCC.
The third inequality is the application of Lemma 2 in Ref.~\cite{TGW14} by regarding ${\cal H}_{{\cal P}_A }^{(i-1)'|{{\bm k}_{i-1}}}$ as system $A$, 
$\tilde{{\cal H}}_{Y|{\bm k}_{i-1}}$ as system $B_1$,
${\cal H}_{E_{{\bm k}_{i-1}}}$ as system $E_1$,
$ {\cal H}_{{\cal P}_B\setminus (Y|{\bm k}_{i-1})}^{(i-1)'|{{\bm k}_{i-1}}} \otimes {\cal H}_{Y|{\bm k}_{i-1}}^{(i-1)|{{\bm k}_{i-2}}}$ as system $B_2$,
and ${\cal H}_{R_{{\bm k}_{i-1}}}$ as system $E_2$.
The final inequality follows from the definition \cite{TGW14} of the squashed entanglement of a quantum channel.
The same inequality is derived if we start by assuming  $X|{{\bm k}_{i-1}} \in {\cal P}_B$ and $Y|{{\bm k}_{i-1}} \in {\cal P}_A$.

Therefore, using Eqs.~(\ref{eq:a1}) and (\ref{eq:a2}) recursively and the fact that $\hat{\rho}_1^{ABC^1C^2\ldots C^n}$ is separable, we obtain
\begin{align}
\sum_{{\bm k_l} } p_{{\bm k}_l} E_{\rm sq}^{{\cal H}_A^{l|{{\bm k}_{l-1}}}:{\cal H}_B^{l|{{\bm k}_{l-1}}}} (\hat{\rho}_{{\bm k}_l}^{AB}) =&  \sum_{{\bm k_{l-1}} } p_{{\bm k}_{l-1}} \sum_{k_l } p_{k_l|{\bm k}_{l-1} } E_{\rm sq}^{{\cal H}_A^{l|{{\bm k}_{l-1}}}:{\cal H}_B^{i|{{\bm k}_{l-1}}}} (\hat{\rho}_{{\bm k}_l}^{AB})  \\
\le&  \sum_{{\bm k_{l-1}} \in K_{\rm out|{\cal P}_A} } p_{{\bm k}_{l-1}}  E_{\rm sq} ( {\cal N}^{\bar{{\cal H}}_{X|{{\bm k}_{l-1}}} \to \tilde{{\cal H}}_{Y|{{\bm k}_{l-1}}} })  \nonumber  \\
&+\sum_{{\bm k_{l-1}} } p_{{\bm k}_{l-1}}  E_{\rm sq}^{{\cal H}_{{\cal P}_A }^{(l-1)|{{\bm k}_{l-2}}}  :{\cal H}_{{\cal P}_B}^{(l-1)|{{\bm k}_{l-2}}} }  (\hat{\rho}_{{\bm k}_{l-1}}^{ABC^1C^2\ldots C^n} ) \\
\le&  \sum_{i=1}^l \sum_{{\bm k_{i-1}} \in K_{\rm out|{\cal P}_A}} p_{{\bm k}_{i-1}}  E_{\rm sq} ( {\cal N}^{\bar{{\cal H}}_{X|{{\bm k}_{i-1}}} \to \tilde{{\cal H}}_{Y|{{\bm k}_{i-1}}} })   .
\end{align}
Combined with Eq.~(\ref{eq:proof1}), this concludes
\begin{align}
\sum_{{\bm k_l} } p_{{\bm k}_l}  \log d_{{\bm k}_l} \le  \frac{1}{1-16\sqrt{\epsilon}} \left(\sum_{i=1}^l \sum_{{\bm k_{i-1}} \in K_{\rm out|{\cal P}_A}} p_{{\bm k}_{i-1}}  E_{\rm sq} ( {\cal N}^{\bar{{\cal H}}_{X|{{\bm k}_{i-1}}} \to \tilde{{\cal H}}_{Y|{{\bm k}_{i-1}}} })  + 
4 h(2\sqrt{\epsilon}) \right). \label{eq:main}
\end{align}
This is equivalent to Eq.~(1).


\begin{thebibliography}{99}
\bibitem{note1}
Note that the bound based on the relative entropy of
entanglement in papers \cite{PL15,PLO15,P16} can be used on channels for which
the teleportation correction operations suitably commute with the
channel. This is a feature held by Pauli channels (e.g., depolarizing
and dephasing channels), erasure channels and Gaussian channels. By contrast, amplitude damping channels are not stretchable \cite{PL15}.
\end{thebibliography}
\end{document}